\documentclass[10pt,conference]{IEEEtran}
\usepackage{epsfig,amsmath,amssymb,amsfonts,amstext,mathrsfs}
\usepackage{latexsym,graphics,epsf,epsfig,psfrag}

\newtheorem{corollary}{\textbf{Corollary}}

\newtheorem{theorem}{\textbf{Theorem}}

\newtheorem{remark}{\textbf{Remark}}

\newcommand{\spp}{\hspace{5mm}}

\newcommand{\mE}{\mathrm{E}}

\newcommand{\cW}{\mathcal{W}}

\newcommand{\cC}{\mathcal{C}}

\newcommand{\cP}{\mathcal{P}}

\newcommand{\uh}{\underline{h}}

\newcommand{\up}{\underline{p}}

\newcommand{\hw}{\hat{w}}

\newcommand{\baralpha}{\bar{\alpha}}

\DeclareMathAlphabet{\matheuf}{U}{euf}{m}{n}

\IEEEoverridecommandlockouts

\begin{document}

\title{Secrecy Capacity Region of Fading Broadcast Channels
\thanks{The research was supported by the National Science Foundation under
Grant Nos.\ ANI-03-38807 and CNS-06-25637.} }
\author{
\authorblockN{Yingbin Liang}
\authorblockA{Dept. of Electrical Engineering \\
Princeton University \\
Princeton, NJ 08544 \\
yingbinl@princeton.edu} \and
\authorblockN{H. Vincent Poor}
\authorblockA{Dept. of Electrical Engineering \\
Princeton University \\
Princeton, NJ 08544 \\
poor@princeton.edu} \and
\authorblockN{Shlomo Shamai (Shitz)}
\authorblockA{Dept. of Electrical Engineering  \\
Technion-Israel Institute of Technology \\
Technion City, Haifa 32000, Israel \\
sshlomo@ee.technion.ac.il} }

\maketitle

\thispagestyle{plain}

\begin{abstract}
The fading broadcast channel with confidential messages (BCC) is
investigated, where a source node has common information for two
receivers (receivers 1 and 2), and has confidential information
intended only for receiver 1. The confidential information needs
to be kept as secret as possible from receiver 2. The broadcast
channel from the source node to receivers 1 and 2 is corrupted by
multiplicative fading gain coefficients in addition to additive
Gaussian noise terms. The channel state information (CSI) is
assumed to be known at both the transmitter and the receivers. The
secrecy capacity region is first established for the parallel
Gaussian BCC, and the optimal source power allocations that
achieve the boundary of the secrecy capacity region are derived.
In particular, the secrecy capacity region is established for the
Gaussian case of the
Csisz$\acute{\text{a}}$r-K$\ddot{\text{o}}$rner BCC model. The
secrecy capacity results are then applied to give the ergodic
secrecy capacity region for the fading BCC.
\end{abstract}

\section{Introduction}

The wire-tap channel models a communication system in which a
source node wishes to transmit confidential information to a
destination node and wishes to keep a wire-tapper as ignorant of
this information as possible. This channel was introduced by Wyner
in \cite{Wyner75}, where the secrecy capacity was given. The
secrecy capacity of the Gaussian wire-tap channel was given in
\cite{Leung78}. The wire-tap channel was considered recently for
fading and multiple antenna channels in \cite{Parada05,Barros06}.
A more general model of the wire-tap channel was studied by
Csisz$\acute{\text{a}}$r and K$\ddot{\text{o}}$rner in
\cite{Csiszar78}, where the source node also has a common message
for both receivers in addition to the confidential message for
only one receiver. This channel is regarded as the broadcast
channel with confidential messages (BCC). The
capacity-equivocation region and the secrecy capacity region of
the discrete memoryless BCC were characterized in
\cite{Csiszar78}. The BCC was further studied recently in
\cite{Liu06}, where the source node transmits two confidential
message sets for two receivers, respectively.

In this paper, we investigate the fading BCC, which is based on
the BCC studied in \cite{Csiszar78} with the channels from the
source node to receivers 1 and 2 corrupted by multiplicative
fading gain coefficients in addition to additive Gaussian noise
terms. We assume that the channel state information (CSI) is known
at both the transmitter and the receivers. The CSI at the
transmitter (the source node) can be realized by reliable feedback
from the two receivers, who are supposed to receive information
from the source node.

The fading BCC we study in this paper relates to or generalizes a
few channels that have been previously studied in the literature.
Compared to the fading broadcast channel studied in
\cite{Hugh95,Tse97,Li01a,Li01,Jindal04}, the fading BCC requires a
secrecy constraint that the confidential information for one
receiver must be as secret as possible from the other receiver.
The fading BCC includes the fading wire-tap channel studied in
\cite{Liang06allerton_a,Li06} and \cite{Gopa06} (full CSI case) as
a special case, because the fading BCC assumes that the source
node has a common message for both receivers in addition to the
confidential message for receiver 1. The fading BCC also includes
the parallel Gaussian wire-tap channel studied in \cite{Yama91}
(the case where wire-tappers cooperate) as a special case for the
same reason as above and also because a power constraint is
assumed for each subchannel in \cite{Yama91}.

In this paper, we first study the parallel Gaussian BCC, which
serves as a basic model that includes the fading BCC as a special
case. We show that the secrecy capacity region of the parallel
Gaussian BCC is a union over the rate regions achieved by all
source power allocations (among the parallel subchannels).
Moreover, we derive the optimal power allocations that achieve the
boundary of the secrecy capacity region and hence completely
characterize this region. In particular, we establish the secrecy
capacity region of the Gaussian case of the
Csisz$\acute{\text{a}}$r-K$\ddot{\text{o}}$rner BCC model.

We then apply our results to study the fading BCC, which can be
viewed as the parallel Gaussian BCC with each fading state
corresponding to one subchannel. Thus, the secrecy capacity region
of the parallel Gaussian BCC applies to the fading BCC. In
particular, since the source node knows the CSI, it can
dynamically change its transmission power with channel state
realization to achieve the boundary of the secrecy capacity
region.

In this paper, we use $X_{[1,L]}$ to indicate a group of variables
$(X_1,X_2,\ldots,X_L)$, and use $X_{[1,L]}^n$ to indicate a group
of vectors $(X_1^n,X_2^n,\ldots,X_L^n)$, where $X_l^n$ indicates
the vector $(X_{l1},X_{l2},\ldots,X_{ln})$. Throughout the paper,
the logarithmic function is to the base $2$.

The paper is organized as follows. We first study the parallel
Gaussian BCC. We then study the fading BCC and demonstrate our
results with numerical examples. We conclude the paper with a few
remarks.


\section{Parallel Gaussian BCCs}\label{sec:gau}

\begin{figure}
\begin{center}
\includegraphics[width=8cm,clip]{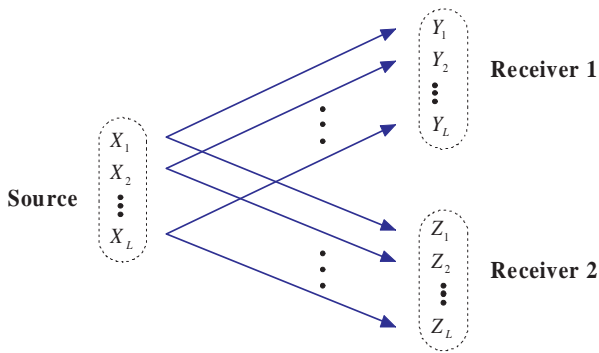}
\caption{Parallel BCC} \label{fig:bcc_para}
\end{center}
\end{figure}

We consider the parallel Gaussian BCC with $L$ independent
subchannels (see Fig.~\ref{fig:bcc_para}), where there are one
source node and two receivers. As in the BCC, the source node
wants to transmit common information to both receivers and
confidential information to receiver 1. Moreover, the source node
wishes to keep the confidential information to be as secret as
possible from receiver 2.

For each subchannel, outputs at receivers 1 and 2 are corrupted by
additive Gaussian noise terms. The channel input-output
relationship is given by
\begin{equation}\label{eq:gaumodel}
Y_{li}=X_{li}+W_{li}, \spp Z_{li}=X_{li}+V_{li}, \spp \text{for
}\; l=1,\ldots,L
\end{equation}
where $i$ is the time index. For $l=1,\ldots,L$, the noise
processes $\{W_{li}\}$ and $\{V_{li}\}$ are independent
identically distributed (i.i.d.) with the components being
Gaussian random variables with the variances $\mu_l^2$ and
$\nu_l^2$, respectively. We assume $\mu_l^2 < \nu_l^2$ for $l\in
A$ and $\mu_l^2 \ge \nu_l^2$ for $l \in A^c$. The channel input
sequence $X_{[1,L]}^n$ is subject to the average power constraints
$P$, i.e.,
\begin{equation}\label{eq:power}
\frac{1}{n}\sum_{i=1}^n \sum_{l=1}^L \mE \left[X_{li}^2\right]
\leq P.
\end{equation}

A $\left( 2^{nR_0},2^{nR_1}, n \right)$ code consists of the
following:
\begin{list}{$\bullet$}{\topsep=0ex \leftmargin=5mm
\rightmargin=2mm \itemsep=0mm}

\item Two message sets: $\cW_0=\{1,2,\ldots,2^{nR_0}\}$ and
$\cW_1=\{1,2,\ldots,2^{nR_1}\}$ with the messages $W_0$ and $W_1$
uniformly distributed over the sets $\cW_0$ and $\cW_1$,
respectively;

\item One (stochastic) encoder at the source node that maps each
message pair $(w_0,w_1) \in (\cW_0,\cW_1) $ to a codeword
$x_{[1,L]}^n $;

\item Two decoders: one at receiver 1 that maps a received
sequence $y_{[1,L]}^n $ to a message pair $(\hw^{(1)}_0,\hw_1) \in
\left(\cW_0,\cW_1 \right)$; the other at receiver 2 that maps a
received sequence $z_{[1,L]}^n $ to a message $\hw^{(2)}_0 \in
\cW_0$.
\end{list}

The secrecy level of the confidential message $W_1$ achieved at
receiver 2 is measured by the following {\em equivocation rate}:
\begin{equation}
\frac{1}{n} H\left(W_1 \Big|Z_{[1,L]}^n\right).
\end{equation}

A rate-equivocation triple $(R_0,R_1,R_e)$ is {\em achievable} if
there exists a sequence of $\left(2^{nR_0},2^{nR_1}, n \right)$
codes with the average probability of error goes to zero as $n$
goes to infinity and with the equivocation rate $R_e$ satisfying
\begin{equation}
R_e \leq \lim_{n\rightarrow \infty} \frac{1}{n} H\left(W_1
\Big|Z_{[1,L]}^n \right).
\end{equation}

In this paper, we focus on the case in which perfect secrecy is
achieved, i.e., receiver 2 does not obtain any information about
the message $W_1$. This happens if $R_e=R_1$. The {\em secrecy
capacity region} $\cC_s$ is defined to be the set that includes
all $(R_0,R_1)$ such that $(R_0,R_1,R_e=R_1)$ is achievable, i.e.,
\begin{equation}
\cC_s=\Big\{(R_0,R_1): (R_0,R_1,R_e=R_1) \text{ is achievable}
\Big\}.
\end{equation}

For the parallel Gaussian BCC, we characterize the secrecy
capacity region in the following Theorems \ref{th:gausec} and
\ref{th:gauoptsl}.
\begin{theorem}\label{th:gausec}
The secrecy capacity region of the parallel Gaussian BCC is
\begin{equation}\label{eq:gausec}
\begin{split}
& \cC^{g}_s =\bigcup_{\up \in \cP} \\
& \left\{
\begin{array}{l}
(R_0,R_1): \\
 R_0 \leq \min \\
\Bigg\{ \displaystyle \sum\limits_{l\in A}
\frac{1}{2}\log\left(1+\frac{p_{l0}}{\mu_l^2+p_{l1}}\right)
+\sum\limits_{l\in A^c} \frac{1}{2}\log\left(1+\frac{p_{l0}}{\mu_l^2}\right), \\
\displaystyle \hspace{3mm} \sum\limits_{l\in A}
\frac{1}{2}\log\left(1+\frac{p_{l0}}{\nu_l^2+p_{l1}}\right)
+\sum\limits_{l\in A^c} \frac{1}{2}\log\left(1+\frac{p_{l0}}{\nu_l^2}\right) \Bigg\} \\
\displaystyle R_1 \leq \sum\limits_{l \in A} \left[
\frac{1}{2}\log\left(1+\frac{p_{l1}}{\mu_l^2}\right)
-\frac{1}{2}\log\left(1+\frac{p_{l1}}{\nu_l^2}\right)\right]\\
\end{array} \right\}
\end{split}
\end{equation}
where $\up$ is the power allocation vector, which consists of
$(p_{l0}, p_{l1})$ for $l\in A$ and $p_{l0}$ for $l \in A^c$ as
components. The set $\cP$ includes all power allocation vectors
$\up$ that satisfy the power constraint \eqref{eq:power}, i.e.,
\begin{equation}\label{eq:uppower}
\cP:=\left\{ \up: \sum_{l\in A} [p_{l0}+p_{l1}]+\sum_{l\in A^c}
p_{l0} \leq P \right\}.
\end{equation}
\end{theorem}
\begin{proof}
The achievability proof uses the following scheme. For $l \in A$,
the source node transmits both common and confidential messages
using the superposition encoding, and $p_{l0}$ and $p_{l1}$
indicate the powers allocated to transmit the common and private
messages, respectively. For $l \in A^c$, the source node transmits
only the common message, and $p_{l0}$ indicates the power to
transmit the common message. The converse proof involves clever
use of the entropy power inequality. Details of the proof can be
found in \cite{Liang06novit}.
\end{proof}

In particular, the converse proof for the parallel Gaussian BCC
also gives the converse proof for the Gaussian BCC ($L=1$), and
hence establishes the following secrecy capacity region for the
Gaussian case of the
Csisz$\acute{\text{a}}$r-K$\ddot{\text{o}}$rner BCC model.
\begin{corollary}\label{cor:cksec}
The secrecy capacity region of the Gaussian BCC is
\begin{equation}\label{eq:cksec}
\begin{split}
& \cC_s =\bigcup_{0 \leq \beta \leq 1} \\
& \left\{
\begin{array}{l}
(R_0,R_1): \\
\displaystyle R_0 \leq \min \Bigg\{ \frac{1}{2}\log\left(1+
\frac{(1-\beta)P}{\mu^2+\beta P}\right), \\
\displaystyle  \hspace{2cm} \frac{1}{2}\log\left(1+
\frac{(1-\beta)P}{\nu^2+\beta P}\right) \Bigg\}\\
\displaystyle R_1 \leq \left[\frac{1}{2}\log\left(1+\frac{\beta
P}{\mu^2}\right) -\frac{1}{2}\log\left(1+\frac{\beta
P}{\nu^2}\right)\right]^+
\end{array} \right\}
\end{split}
\end{equation}
where $(x)^+=x$ if $x > 0$ and $(x)^+=0$ if $x \leq 0$.
\end{corollary}

Note that the secrecy capacity region of the parallel Gaussian BCC
given in \eqref{eq:gausec} is convex. Hence the boundary of this
region can be characterized as follows. For every point
$(R^*_0,R^*_1)$ on the boundary, there exist $\gamma_0
>0$ and $\gamma_1>0$ such that $(R^*_0,R^*_1)$ is the solution to
the following problem
\begin{equation}
\max_{(R_0,R_1) \in \cC_s^{g}} \Big[\gamma_0 R_0 + \gamma_1
R_1\Big].
\end{equation}
Therefore, the power allocation $\up^*$ that achieves the boundary
point $(R^*_0,R^*_1)$ is the solution to the following problem
\begin{equation}\label{eq:gauopt}
\begin{split}
\max_{\up \in \cP} & \Big[\gamma_0 R_0(\up) + \gamma_1 R_1(\up)\Big] \\
& = \max_{\up \in \cP} \Big[\gamma_0 \min \left\{
R_{01}(\up),R_{02}(\up) \right\} + \gamma_1 R_1(\up)\Big]
\end{split}
\end{equation}
where $R_0(\up)$ and $R_1(\up)$ indicate the bounds on $R_0$ and
$R_1$ in \eqref{eq:gausec}. We further define $R_{01}(\up)$ and
$R_{02}(\up)$ to be the two terms over which the minimization in
$R_0(\up)$ is taken, i.e., $R_0(\up)=\min
\{R_{01}(\up),R_{02}(\up) \}$. The solution to \eqref{eq:gauopt}
is given in the following theorem. The proof can be found in
\cite{Liang06novit} and is omitted here due to space limitations.
\begin{theorem}\label{th:gauoptsl}
The optimal power allocation vector $\up^*$ that solves
\eqref{eq:gauopt} and hence achieves the boundary of the secrecy
capacity region of the parallel Gaussian BCC has one of the
following three forms.

\begin{small}
Case 1: $\up^*=\up^{(1)}$ if the following $\up^{(1)}$ satisfies
$R_{01}\left(\up^{(1)}\right) < R_{02}\left(\up^{(1)}\right)$.
\begin{equation*}
\begin{split}
& \text{For} \;\; l \in A, \;\; \text{if} \;\;
\frac{\gamma_1}{\gamma_0}
> \frac{\nu_l^2}{\nu_l^2-\mu_l^2}, \\
& p^{(1)}_{l0}=\left(\frac{\gamma_0}{2\lambda\ln
2}-\left(\frac{\gamma_1}{\gamma_0}-1
\right)(\nu_l^2-\mu_l^2)\right)^+,
\\
& p^{(1)}_{l1}= \\
& \Bigg(\min \Bigg\{\frac{1}{2}\sqrt{(\nu_l^2-\mu_l^2)
\left(\nu_l^2-\mu_l^2+\frac{2\gamma_1}{\lambda\ln
2}\right)}-\frac{1}{2}\left( \mu_l^2+\nu_l^2 \right), \\
& \hspace{1.3cm}\frac{\gamma_1}{\gamma_0}(\nu_l^2-\mu_l^2)-\nu_l^2
\Bigg\}\Bigg)^+, \\
& \text{if}\;\; \frac{\gamma_1}{\gamma_0} \leq
\frac{\nu_l^2}{\nu_l^2-\mu_l^2}, \;\;
p^{(1)}_{l0}=\left(\frac{\gamma_0}{2\lambda\ln
2}-\mu_l^2\right)^+, \spp p^{(1)}_{l1}=0; \\
& \text{For}\;\; l \in A^c, \;\; p^{(1)}_{l0}= \left(
\frac{\gamma_0}{2\lambda\ln 2}-\mu_l^2 \right)^+ \hspace{2.4cm}
\end{split}
\end{equation*}
where $\lambda$ is chosen to satisfy the power constraint
\begin{equation}\label{eq:gaupower}
\sum_{l\in A} [p_{l0}+p_{l1}]+\sum_{l\in A^c} p_{l0} \leq P.
\end{equation}

Case 2: $\up^*=\up^{(2)}$ if the following $\up^{(2)}$ satisfies
$R_{01}\left(\up^{(2)}\right) > R_{02}\left(\up^{(2)}\right)$.
\begin{equation*}\label{eq:gauopt2}
\begin{split}
& \text{For} \;\; l \in A, \;\; \text{if} \;\;
\frac{\gamma_1}{\gamma_0}
> \frac{\mu_l^2}{\nu_l^2-\mu_l^2}, \\
& p^{(2)}_{l0}=\left(\frac{\gamma_0}{2\lambda\ln
2}-\left(\frac{\gamma_1}{\gamma_0}+1
\right)(\nu_l^2-\mu_l^2)\right)^+,
\\
& p^{(2)}_{l1}= \\
& \Bigg(\min \Bigg\{\frac{1}{2}\sqrt{(\nu_l^2-\mu_l^2)
\left(\nu_l^2-\mu_l^2+\frac{2\gamma_1}{\lambda\ln 2}\right)}
-\frac{1}{2}\left( \mu_l^2+\nu_l^2 \right), \\
& \hspace{1.3cm}
\frac{\gamma_1}{\gamma_0}(\nu_l^2-\mu_l^2)-\mu_l^2 \Bigg\}\Bigg)^+, \\
& \text{if}\;\; \frac{\gamma_1}{\gamma_0} \leq
\frac{\mu_l^2}{\nu_l^2-\mu_l^2},
\;\;\; p^{(2)}_{l0}=\left(\frac{\gamma_0}{2\lambda\ln 2}-\nu_l^2\right)^+, \spp p^{(1)}_{l1}=0; \\
& \text{For}\;\; l \in A^c, \;\; p^{(2)}_{l0}= \left(
\frac{\gamma_0}{2\lambda\ln 2}-\nu_l^2 \right)^+
\end{split}
\end{equation*}
where $\lambda$ is chosen to satisfy \eqref{eq:gaupower}.

Case 3: $\up^*=\up^{(\alpha)}$ if there exists $0 \leq \alpha \leq
1$ such that the following $\up^{(\alpha)}$ satisfies
$R_{01}\left(\up^{(\alpha)}\right)=
R_{02}\left(\up^{(\alpha)}\right)$.
\begin{equation*}
\begin{split}
& \text{For} \;\; l \in A, \;\; \text{if} \;\;
\frac{\gamma_1}{\gamma_0}
> \frac{\alpha\nu_l^2+\baralpha\mu_l^2}{\nu_l^2-\mu_l^2}, \\
&
p^{(\alpha)}_{l0}=\Bigg(\frac{1}{2}\sqrt{\left(\nu_l^2-\mu_l^2-\frac{\gamma_0}{2\ln
2 \lambda}\right)^2+\frac{2\alpha\gamma_0}{\lambda\ln
2}(\nu_l^2-\mu_l^2)}\\
& \hspace{1.4cm} +\frac{\gamma_0}{4\ln 2
\lambda}-\left(\frac{\gamma_1}{\gamma_0}-\alpha+\frac{1}{2}
\right)(\nu_l^2-\mu_l^2)\Bigg)^+,
\\
& p^{(\alpha)}_{l1}= \\
& \Bigg(\min \Bigg\{\frac{1}{2}\sqrt{(\nu_l^2-\mu_l^2)
\left(\nu_l^2-\mu_l^2+\frac{2\gamma_1}{\lambda\ln 2}\right)}
-\frac{1}{2}\left( \mu_l^2+\nu_l^2 \right), \\
& \hspace{1.3cm}
\frac{\gamma_1}{\gamma_0}(\nu_l^2-\mu_l^2)-(\alpha\nu_l^2+\baralpha\mu_l^2)
\Bigg\}\Bigg)^+, \\
 & \text{if}\;\; \frac{\gamma_1}{\gamma_0} \leq
\frac{\alpha\nu_l^2+\baralpha\mu_l^2}{\nu_l^2-\mu_l^2}, \\
&
p^{(\alpha)}_{l0}=\Bigg(\frac{1}{2}\sqrt{\left(\nu_l^2-\mu_l^2-\frac{\gamma_0}{2\ln
2 \lambda}\right)^2+\frac{2\alpha\gamma_0}{\lambda\ln
2}(\nu_l^2-\mu_l^2)} \\
&
\hspace{1.3cm}-\frac{1}{2}\left(\mu_l^2+\nu_l^2-\frac{\gamma_0}{2\ln
2\lambda} \right) \Bigg)^+, \\
& p^{(\alpha)}_{l1}=0; \\
& \text{For}\;\; l \in A^c, \\
& p^{(\alpha)}_{l0}=
\Bigg(\frac{1}{2}\sqrt{\left(\nu_l^2-\mu_l^2-\frac{\gamma_0}{2\ln
2 \lambda}\right)^2+\frac{2\alpha\gamma_0}{\lambda\ln
2}(\nu_l^2-\mu_l^2)}
\\
&
\hspace{1.3cm}-\frac{1}{2}\left(\mu_l^2+\nu_l^2-\frac{\gamma_0}{2\ln
2\lambda} \right) \Bigg)^+
\end{split}
\end{equation*}
where $\lambda$ is chosen to satisfy \eqref{eq:gaupower}.
\end{small}
\end{theorem}

Based on Theorem \ref{th:gauoptsl}, we provide the following
algorithm to search the optimal $\up^*$.
\begin{center}
\begin{small}
Algorithm to search $\up^*$ that solves \eqref{eq:gauopt}
\begin{tabular}{ll}
\hline \\
 Step 1. & \small Find $\up^{(1)}$ given in Case 1 in Theorem \ref{th:gauoptsl}. \\
& \small If $R_{01}\left(\up^{(1)}\right) <
R_{02}\left(\up^{(1)}\right)$, then $\up^*=\up^{(1)}$ and finish. \\
& \small Otherwise, go to Step 2. \\
Step 2. & \small Find $\up^{(2)}$ given in Case 2 in Theorem \ref{th:gauoptsl}. \\
& \small If $R_{01}\left(\up^{(2)}\right)>
R_{02}\left(\up^{(2)}\right)$,
then $\up^*=\up^{(2)}$ and finish. \\
& \small Otherwise, go to Step 3. \\
Step 3. & For a given $\alpha$, find $\up^{(\alpha)}$ given in
Case 3 in Theorem \ref{th:gauoptsl}. \\
& \small Search over $0 \leq \alpha \leq 1$ to find $\alpha$ that
satisfies \\
& $R_{01}\left(\up^{(\alpha)}\right)=
R_{02}\left(\up^{(\alpha)}\right)$. Then $\up^*=\up^{(\alpha)}$ and finish. \\
\hline
\end{tabular}
\end{small}
\end{center}

A numerical example that demonstrates power allocations following
from three cases is given in Section \ref{sec:ergodic}.
\section{Fading BCCs}\label{sec:ergodic}

In this section, we study the fading BCC, where the channel
input-output relationship is given by
\begin{equation}\label{eq:fdmodel}
Y_i = h_{1i} X_i+W_i, \spp Z_i = h_{2i} X_i+V_i
\end{equation}
where $i$ is the time index. The channel gain coefficients
$h_{1i}$ and $h_{2i}$ are proper complex random variables. We
define $\uh_i:=(h_{1i},h_{2i})$, and assume $\{\uh_i\}$ is a
stationary and ergodic vector random process. The noise processes
$\{W_i\}$ and $\{V_i\}$ are i.i.d.\ proper complex Gaussian with
$W_i$ and $V_i$ having variances $\mu^2$ and $\nu^2$,
respectively. The input sequence $\{X_i \}$ is subject to the
average power constraint $P$, i.e., $\frac{1}{n}\sum_{i=1}^n
\mE\big[ X^2_i \big]\leq P$.

We assume that the channel state information (i.e., the
realization of $\uh_i$) is known at both the transmitter and the
receivers instantaneously. The fading BCC can be viewed as a
parallel Gaussian BCC with each fading state corresponding to one
subchannel. Thus, the following secrecy capacity region of the
fading BCC follows from Theorem \ref{th:gausec}.
\begin{corollary}\label{cor:fdsec}
The secrecy capacity region of the fading BCC is
\begin{equation}\label{eq:fdsec}
\begin{split}
& \cC_s =\bigcup_{(p_0(\uh),p_1(\uh)) \in \cP} \\
& \left\{
\begin{array}{l}
(R_0,R_1): \\
\displaystyle R_0 \leq \min \Bigg\{ \mE_{\uh \in
A}\log\left(1+\frac{p_0(\uh)|h_1|^2}{\mu^2+p_1(\uh)|h_1|^2}\right)\\
\displaystyle \hspace{2cm} +\mE_{\uh \in A^c}
\log\left(1+\frac{p_0(\uh)|h_1|^2}{\mu^2}\right), \\
\displaystyle \hspace{2.cm} \mE_{\uh \in A}
\log\left(1+\frac{p_0(\uh)|h_2|^2}{\nu^2+p_1(\uh)|h_2|^2}\right)
\\
\displaystyle \hspace{2cm} +\mE_{\uh
\in A^c}\log\left(1+\frac{p_0(\uh)|h_2|^2}{\nu^2}\right) \Bigg\} \\
\displaystyle R_1 \leq \mE_{\uh \in A} \Bigg[
\log\left(1+\frac{p_1(\uh)|h_1|^2}{\mu^2}\right) \\
\displaystyle \hspace{2cm}-\log\left(1+\frac{p_1(\uh)|h_2|^2}{\nu^2}\right)\Bigg]\\
\end{array} \right\}.
\end{split}
\end{equation}
where $A:= \Big\{\uh: \frac{|h_1|^2}{\mu^2} >
\frac{|h_2|^2}{\nu^2} \Big\}$. The random vector $\uh=(h_1,h_2)$
has the same distribution as the marginal distribution of the
process $\{\uh_i\}$ at one time instant. The functions $p_0(\uh)$
and $p_1(\uh)$ indicate the source powers allocated to transmit
the common and confidential messages, respectively. The set $\cP$
is defined as
\begin{equation}\label{eq:fdpowerset}
\cP=\Big\{(p_0(\uh),p_1(\uh)): \mE_A
\left[p_0(\uh)+p_1(\uh)\right]+\mE_{A^c} [p_0(\uh)] \leq P \Big\}.
\end{equation}
\end{corollary}

From the bound on $R_1$ in \eqref{eq:fdsec}, it can be seen that
as long as $A$ is not a zero probability event, positive secrecy
rate can be achieved. Since fading introduces more randomness to
the channel, it is more likely that the channel from the source
node to receiver 1 is better than the channel from the source node
to receiver 2 for some channel states, and hence positive secrecy
capacity can be achieved by exploiting these channel states.

Since the source node is assumed to know the channel state
information, it can allocate its power according to the
instantaneous channel realization to achieve the best performance,
i.e., the boundary of the secrecy capacity region. Such optimal
power allocations can be derived from Theorem \ref{th:gauoptsl}.
The details can be found in \cite{Liang06novit}.
\begin{remark}
If the source node does not have common messages for both
receivers, and only has confidential messages for receiver 1, the
fading BCC becomes the fading wire-tap channel. For this channel,
Corollary \ref{cor:fdsec} and Theorem \ref{th:gauoptsl} give the
secrecy capacity and the optimal source power allocation obtained
in \cite{Liang06allerton_a,Li06} and \cite{Gopa06} (full CSI
case).
\end{remark}

\begin{figure}
\begin{center}
\includegraphics[width=7.5cm]{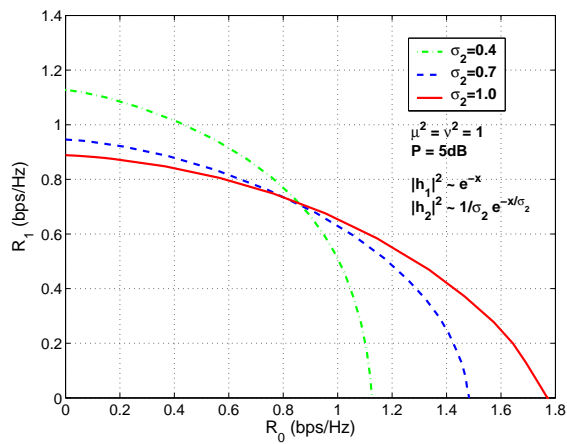}
\caption{Secrecy capacity regions for Rayleigh fading BCCs}
\label{fig:region_gauss}
\end{center}
\end{figure}

We now provide numerical results for the fading BCC. We consider
the Rayleigh fading BCC, where $h_1$ and $h_2$ are zero mean
proper complex Gaussian random variables. Hence $|h_1|^2$ and
$|h_2|^2$ are exponentially distributed with parameters $\sigma_1$
and $\sigma_2$. We assume the source power $P=5$ dB, and fix
$\sigma_1=1$. In Fig.~\ref{fig:region_gauss}, we plot the
boundaries of the secrecy capacity regions corresponding to
$\sigma_2=0.4,0.7,1$, respectively. It can be seen that as
$\sigma_2$ decreases, the secrecy rate $R_1$ of the confidential
message improves, but the rate $R_0$ of the common message
decreases. This fact follows because smaller $\sigma_2$ implies
worse channel from the source node to receiver 2. Thus,
confidential information can be forwarded to receiver 1 at a
larger rate. However, the rate of the common information is
limited by the channel from the source node to receiver 2, and
hence decreases as $\sigma_2$ decreases.

\begin{figure}
\begin{center}
\includegraphics[width=8cm]{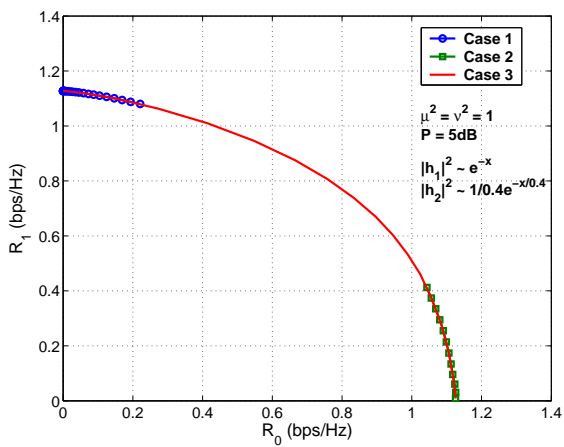}
\caption{Three cases in power allocation optimization to achieve
the boundary of the secrecy capacity region for a Rayleigh fading
BCC} \label{fig:cases}
\end{center}
\end{figure}

For the Rayleigh fading BCC with $\sigma_1=1$ and $\sigma_2=0.4$,
we plot the boundary of the secrecy capacity region in
Fig.~\ref{fig:cases}. The three cases (see Theorem
\ref{th:gauoptsl}) to derive the boundary achieving power
allocations are also indicated with the corresponding boundary
points. It can be seen that the boundary points with large $R_1$
are achieved by the power allocations derived from Case 1, and are
indicated by the line with circle on the graph. The boundary
points with large $R_0$ are achieved by the optimal power
allocations derived from Case 2, and are indicated by the line
with square. Between the boundary points achieved by Case 1 and
Case 2, the boundary points are achieved by the power allocations
derived from Case 3, and are indicated by the plain solid line.

An intuitive reason why the three cases associate with the
boundary points is given as follows. To achieve large secrecy rate
$R_1$, most channel states in the set $A$ where receiver 1 has a
stronger channel than receiver 2 are used to transmit the
confidential message. The common message is hence transmitted
mostly over the channel states in the set $A^c$, over which the
common rate is limited by the channel from the source node to
receiver 1. Thus, power allocation needs to optimize the rate of
this channel, and hence the optimal power allocation follows from
Case 1. To achieve large $R_0$, the common message is forwarded
over the channel states both in $A$ and $A^c$. Since in average
the source node has a much worse channel to receiver 2 than to
receiver 1, the channel from the source node to receiver 2 limits
the common rate. Power allocation now needs to optimize the rate
to receiver 2, and hence follows from Case 2. Between these two
cases, power allocation needs to balance the rates to receivers 1
and 2 and hence follows from Case 3.
\section{Conclusions}

We have established the secrecy capacity region for the parallel
Gaussian BCC, and have characterized the optimal power allocations
that achieve the boundary of this region. An interesting result we
have established is the secrecy capacity region of the Gaussian
case of the Csisz$\acute{\text{a}}$r and K$\ddot{\text{o}}$rner
BCC model.

We have further applied our results to obtain the ergodic secrecy
capacity region for the fading BCC. Our results generalize the
secrecy capacity of the fading wire-tap channel that has been
recently obtained in \cite{Liang06allerton_a}, \cite{Li06} and
\cite{Gopa06} (full CSI case). We have also studied the outage
performance of the fading BCC, the results of which are not
presented in this paper due to space limitations; details can be
found in \cite{Liang06novit}.




\bibliographystyle{IEEEtran}

%

\end{document}